\documentclass{ws-ijbc}
\usepackage{ws-rotating}     % used only when sideways tables/figures are used
\begin{document}

\catchline{}{}{}{}{} % Publisher's Area please ignore

\markboth{DIEGO F. M. OLIVEIRA,MARKO ROBNIK}{SCALING INVARIANCE IN A TIME-DEPENDENT ELLIPTICAL BILLIARD}

\title{SCALING INVARIANCE IN A TIME-DEPENDENT ELLIPTICAL BILLIARD}

\author{DIEGO F. M. OLIVEIRA and MARKO ROBNIK}

\address{{CAMTP - Center For Applied Mathematics and Theoretical Physics - University of Maribor \\ Krekova 2 - SI-2000 - Maribor - Slovenia.}\\
diegofregolente@gmail.com, robnik@uni-mb.si}

\maketitle

\begin{history}
\received{(to be inserted by publisher)}
\end{history}

\begin{abstract}
We study some dynamical properties of a classical time-dependent elliptical billiard. We consider periodically moving boundary 
and collisions between the particle and the boundary are assumed to be elastic. Our results confirm that although the static elliptical billiard is an integrable system, 
after to introduce time-dependent 
perturbation on the boundary
the unlimited energy growth is observed. The behaviour of the average velocity is described using scaling arguments.
\end{abstract}

\keywords{Billiards; Fermi acceleration; scaling.}

%\begin{multicols}{2}
\section{Introduction}

The idea of unlimited energy growth (also known as the Fermi acceleration) due to the repeated collisions of particles with a moving wall was introduced by Enrico Fermi \cite{ref1} 
in  1949 as an attempt to explain the acceleration of cosmic rays. He proposed a
very simple model where charged particles could be accelerated by collisions with moving magnetic field structures. His original model was later modified and studied 
considering 
different approaches. Many of them take into account the inclusion of external fields, simplifications, damping coefficients, quantum and relativistic effects.  

One of the most important versions of this problem is the well known Fermi-Ulam Model (FUM). This model consists of a classical point particle 
of mass $m$, bouncing between two rigid walls. One of them is assumed to be fixed while the other one moves according 
to a periodic function. It is important to emphasize that this system has a very rich phase space structure in the sense that, depending on the initial conditions and control 
parameters, 
one can observe invariant spanning curves, chaotic seas and Kolmogorov-Arnold-Moser (KAM) islands. 
Later on Pustylnikov \cite{ref2,ref3} replaced the fixed wall existing in the Fermi-Ulam model by a constant 
gravitational field \cite{ref4,ref5,ref6,ref7,ref8} the so-called bouncer model. Despite the
similarity between the two models, there is a huge difference between them mainly regarding the average velocity of the particle for long time. 
In the Fermi-Ulam model it was rigorously proved by Pustylnikov that the existence of invariant curves in the phase space always prevents the unlimited energy growth \cite{ref8a} although other 
workers provided an evidence for such a conclusion \cite{refz,refx,refw}. On the other hand, for specific
combinations of both control parameters and initial conditions the phenomenon of unlimited energy growth can be observed in the Pustylnikov bouncer model. This surprising result 
was later discussed and
explained by Lichtenberg and Lieberman \cite{ref9,ref10} and can be easily understood by looking at the phase space. The FUM has
a set of invariant spanning curves limiting the size of the chaotic sea (as well as the particle's velocity), but such invariant tori are not observed in the bouncer model and the energy can typically grow  unbounded. 
A natural extension of the one dimensional billiard models are the two-dimensional billiard systems. Basically they are classified (i) integrable, (ii) ergodic and (iii) mixed. In case (i) the phase space consists of invariant 
tori filling the entire phase space and typical examples are the circular and the elliptical billiard whose the integrability in the case of the circle  comes from the angular momentum conservation, and 
the product of the angular momenta with respect to the foci in case of ellipse \cite{ref00,ref000}. 
In case (ii) the time evolution of a single initial condition is enough to fill the phase space and two examples are the Bunimovich stadium \cite{refa} and the Sinai billiard \cite{refb}. 
In case (iii), there is a representative number of billiards that present mixed phase space
structure \cite{refh,refd,refe,reff,refg,refc}. One important property in the mixed phase space is that chaotic seas are generally surrounding Kolmogorov-Arnold-Moser (KAM) islands which are
confined by invariant spanning curves \cite{ref10,ref2,ref3}. In particular such  curves can cross the phase plane and partition it into several separated
portions of the phase space.  One of the main questions about two dimensional time-dependent systems is: Under what conditions the unlimited energy growth will be observed?  In
this sense,  a conjecture was proposed by Loskutov-Ryabov-Akinshin (LRA) \cite{ref11} and later it was proved by Gelfreich and Turaev \cite{ref11a,ref11b}. This conjecture, known as LRA-conjecture, 
states that the existence of a chaotic component
in the phase space with static boundary is a sufficient condition to observe Fermi acceleration when  a perturbation is introduced.
Very recently Leonel and Bunimovich \cite{refhh} extended the conjecture to the existence of a heteroclinic orbit in the phase space instead of the existence of a set with chaotic dynamics.
 Results that corroborate the validity of this conjecture include the time-dependent oval billiard \cite{ref12}, stadium billiard \cite{ref13} and Lorentz gas \cite{ref13a}.
 Lenz et al. \cite{ref14a,ref14,ref15}  observed that after a time-dependent perturbation is introduced the separatrix gives place to a chaotic layer and the particle's velocity can grow unbounded. Initially it 
was proposed that the acceleration exponent was controlled by the driven amplitude of the boundary \cite{ref14a}. Later, when extensive simulations were taken into account, the authors observed that when the number 
of collisions with the boundary is large enough the acceleration exponent is the same independent of the driving amplitude \cite{ref15}. Such result has been confirmed in \cite{ref13b} and it is studied in more details in the 
present paper.

In this paper we revisit the problem of a classical particle confined in a time-dependent closed region of elliptical shape. 
Our main goal is to describe and understand the behaviour of the average velocity as a function of the number of collisions with the boundary using scaling arguments.

The paper is organized as follows. In section \ref{sec2} we describe the necessary details to define the four-dimensional mapping that describes the dynamics of the
system, and our numerical results. Conclusions are drawn in section \ref{sec3}.

\section{The model and the map.}
\label{sec2}

\begin{figure}[t]
%%\vspace*{-1.1cm}
\centerline{\includegraphics[width=0.60\linewidth]{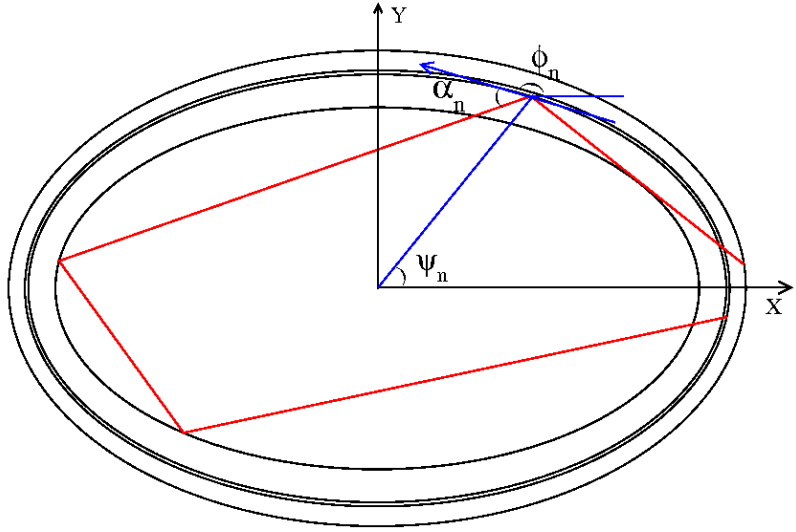}}
\caption{{Illustration of five collision with the time-dependent boundary. 
The corresponding angles that describe the dynamics are also
illustrated.}}
\label{fig1}
\end{figure}

In this section we discuss all the details required for the construction of the mapping. The two-dimensional driven elliptical billiard consists of a classical point particle of mass {\it m} 
confined into a closed region within which it is freely moving and is suffering elastic collisions with the time-dependent boundary.  
We stress that the particle is not affected by any external field  
 and travels freely on a straight line until it reaches the boundary (see Fig. \ref{fig1} ). We describe the dynamics of the system in terms of a 
four-dimensional nonlinear mapping $T(\theta_{n},\alpha_{n},{V}_{n},t_{n})=(\theta_{n+1},
\alpha_{n+1},{V}_{n+1},t_{n+1})$  that gives: the angular position\footnote{ $\theta$ is the canonical parameter of the ellipse (called the eccentric anomaly in astronomy) and is not the angle $\psi$ between the position
vector $(x,y)$ and the x-axis (polar angle). The connection between $\theta$ and the canonical parameter $\psi$ is
$\tan \psi=(B_0/A_0)\tan \theta$. } of the particle $\theta_{n}$; 
the angle that the trajectory of the particle forms with the tangent line at the 
position of the collision $\alpha_{n}$; the absolute value of the particle velocity $V_n=\vert \overrightarrow{V}_{n} \vert$ and the instant of the hit with the boundary $t_n$ \cite{ref12}. 
The index $n$ denotes the $n^{th}$ collision with the moving boundary. 
The Cartesian components of the boundary at the angular position $(\theta_n,t_n)$ are

\begin{eqnarray}
X(\theta_{n},t_n)&=&[A_0+C\sin(t_n)]\cos(\theta_{n})~, \\ 
Y(\theta_{n},t_n)&=&[B_0+C\sin(t_n)]\sin(\theta_{n})~,
\label{eq2}
\end{eqnarray}
where $A_0$ and $B_0$ are constants, thus, at any time $t_n$ we have elliptical shape. The control (oscillation) parameter $0<C<min(A_0,B_0)$ controls the 
amplitude of oscillation and $\theta\in[0,2\pi)$ is a
counterclockwise canonical angle measured with respect to the positive horizontal
axis. The angle between the tangent of the boundary at the position
$(X(\theta_{n}),Y(\theta_{n}))$ measured with respect to the horizontal
line is
\begin{eqnarray}
\phi_n=\arctan\left[Y'(\theta_n) \over X'(\theta_n) \right]~,
\label{eq3}
\end{eqnarray}
where the expressions for both $X'(\theta_n)$ and $Y'(\theta_n)$ are written as
\begin{eqnarray}
X'(\theta_n)&=&{-[A_0+C\sin(t_n)]\sin(\theta_n)}, \\
Y'(\theta_n)&=&{+[B_0+C\sin(t_n)]\cos(\theta_n)}.
\label{eq4}
\end{eqnarray}

\begin{figure}[t]
%\vspace*{-0.8cm}
\centerline{\includegraphics[width=0.60\linewidth]{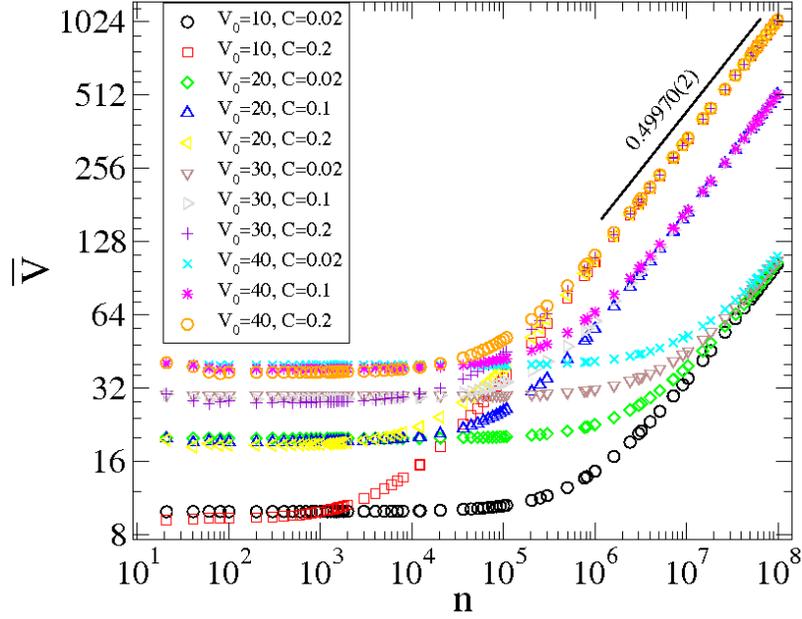}}
\caption{{(Color online) Behaviour of ${\bar V} vs. n$ for different initial
velocities. The control parameters used were $A_0=2$ , $B_0=1$.}}
\label{fig19}

\end{figure}
Since both $\phi_n$ and $\alpha_n$ are already known, the angle between the horizontal axis and the particle's trajectory is $(\phi_n+\alpha_n)$ and the vector velocity of 
the particle is given by
\begin{eqnarray}
\overrightarrow{V}_n=\vert\overrightarrow{V_n}\vert [\cos(\phi_n+\alpha_n)\widehat{i}+\sin(\phi_n+\alpha_n)\widehat{j}]~,
\label{eq5}
\end{eqnarray}
where $\widehat{i}$ and $\widehat{j}$ represent the unit vectors with respect
to the X and Y axis, respectively. This allows us to obtain the  position of the particle as function of time as follows
\begin{eqnarray}
X_p(t)&=&X(\theta_n,t_n)+\vert\overrightarrow{V}_n\vert\cos(\phi_n + \alpha_n)(t-t_n)~,\\
Y_p(t)&=&Y(\theta_n,t_n)+\vert\overrightarrow{V}_n\vert\sin(\phi_n + \alpha_n)(t-t_n)~,
\label{eq6}
\end{eqnarray}
where the index $p$ denotes that such coordinates correspond to the particle. In order to find the angular position $\theta_{n+1}$ at the collision $n+1$, and the time $t_{n+1}$ of the 
next collision, we need to solve numerically the 
expression $R_p(\theta_{n+1},t_{n+1})=R_b(\theta_{n+1},t_{n+1})$ where $R_p(\theta_n,t_n)$ is the distance of the particle measured with respect to the origin of the coordinate 
system, i.e., $R_p(\theta_n,t_n)=\sqrt{X^2_p(t)+Y^2_p(t)}$, and by evaluating the expression
\begin{eqnarray}
t_{n+1}=t_n+ {{\sqrt{[X_p(t_{n+1})-X(\theta_{n},t_{n})]^2+[Y_p(t_{n+1})-Y(\theta_{n},t_{n})]^2}} \over \vert\overrightarrow{V}_n\vert}~.
\label{eq7}
\end{eqnarray}
This is equivalent to solving the two equations $X_p(t_{n+1})=X(\theta_{n+1},t_{n+1})$ and $Y_p(t_{n+1})=Y(\theta_{n+1},t_{n+1})$. In the comoving frame of the boundary,
 where the velocity of the particle for the elastic collision is denoted by $\overrightarrow{V}^\prime$, the following conditions must be satisfied
\begin{eqnarray}
\overrightarrow{V^\prime}_{n+1}.\overrightarrow{T}_{n+1}&=&\overrightarrow{V^\prime}_{n}.\overrightarrow{T}_{n+1}~,\\
\label{eq8}
\overrightarrow{V^\prime}_{n+1}.\overrightarrow{N}_{n+1}&=&-\overrightarrow{V^\prime}_{n}.\overrightarrow{N}_{n+1}~.
\label{eq9}
\end{eqnarray}
At the new angular position, $\theta_{n+1}$, the unitary tangent and normal vectors are
\begin{eqnarray}
\overrightarrow{T}_{n+1}&=&\cos(\phi_{n+1})\widehat{i}+\sin(\phi_{n+1})\widehat{j}~,\\
\overrightarrow{N}_{n+1}&=&-\sin(\phi_{n+1})\widehat{i}+\cos(\phi_{n+1})\widehat{j}~,
\label{eq10}
\end{eqnarray}
and after some algebra we can easily find
\begin{eqnarray}
\overrightarrow{V}_{n+1}.\overrightarrow{T}_{n+1}&=&\mid \overrightarrow{V_n} \mid[\cos(\phi_{n+1}-\phi_n-\alpha_n)],\\
\overrightarrow{V}_{n+1}.\overrightarrow{N}_{n+1}&=&\mid \overrightarrow{V_n}\mid [\sin(\phi_{n+1}-\phi_n-\alpha_n)]+ 2\overrightarrow{V}_{b}.\overrightarrow{N}_{n+1},
\label{eq11}
\end{eqnarray}
where $\overrightarrow{V}_{b}$ is the velocity of the boundary which is written as 
\begin{eqnarray}
\overrightarrow{V}_{b}&=&C\cos(t_{n+1})[ \cos(\theta_{n+1})\widehat{i}+\sin(\theta_{n+1})\widehat{j}]~.
\label{eq12}
\end{eqnarray}
Thus, the absolute value of the velocity just after the collision $(n+1)$ is given by
\begin{eqnarray}
{V}_{n+1}=\mid \overrightarrow{V}_{n+1}\mid=\sqrt{[\overrightarrow{V}_{n+1}.\overrightarrow{T}_{n+1}]^2+[\overrightarrow{V}_{n+1}.\overrightarrow{N}_{n+1}]^2}~,
\label{eq13}
\end{eqnarray} 
and the angle $\alpha_{n+1}$ is
\begin{eqnarray}
\tan \alpha_{n+1}= {\overrightarrow{V}_{n+1}.\overrightarrow{N}_{n+1} \over \overrightarrow{V}_{n+1}.\overrightarrow{T}_{n+1}} ~.
\label{eq14}
\end{eqnarray}

\begin{figure}[t]
%\vspace*{-0.8cm}
\centerline{\includegraphics[width=0.60\linewidth]{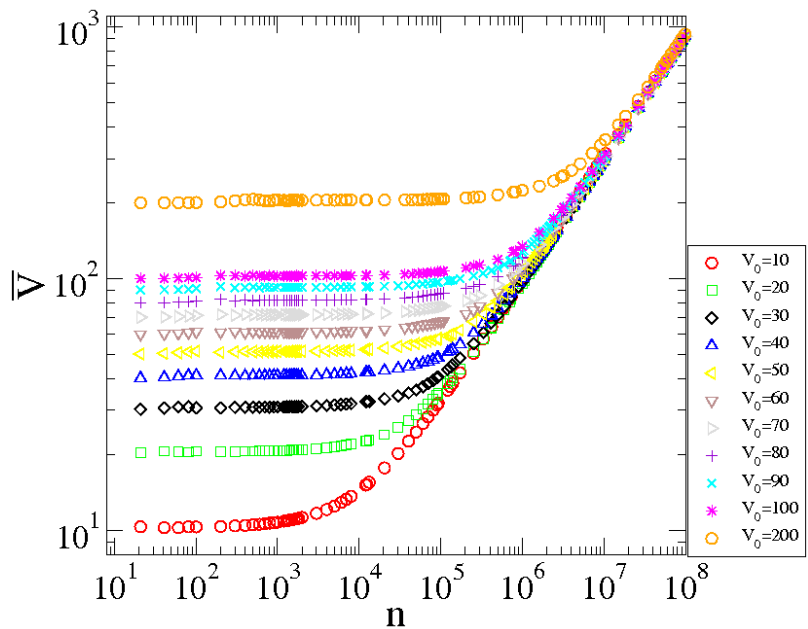}}
\caption{{Behaviour of ${\bar V}\times n$ for different initial
velocities. The control parameters used were $A_0=2$, $B_0=1$, $C=0.2$.}}
\label{fig20a}
\end{figure}

\begin{figure}[t]
%\vspace*{-0.7cm}
\centerline{\includegraphics[width=0.60\linewidth]{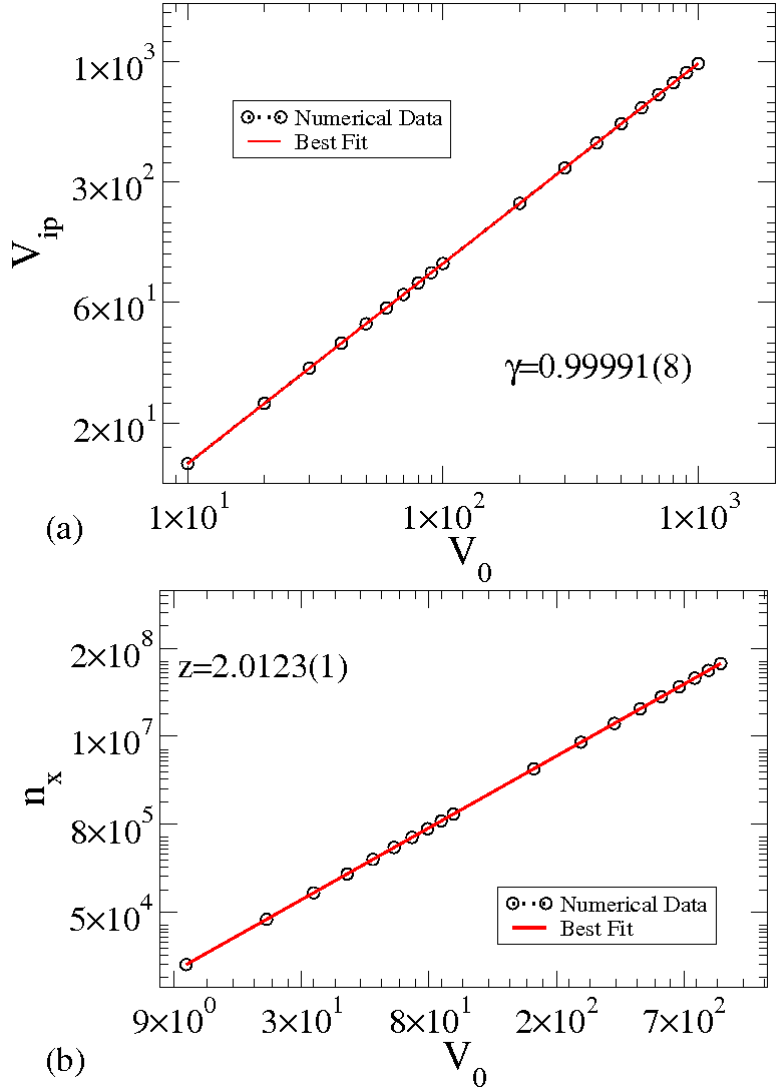}}
\caption{{(a) Plot of $V_{ip} \times V_0$. (b) Behaviour of $n_x$ as
function of $V_0$.}}
\label{fig20}
\end{figure}

\begin{figure}[t]
%%\vspace*{-1.1cm}
\centerline{\includegraphics[width=0.60\linewidth]{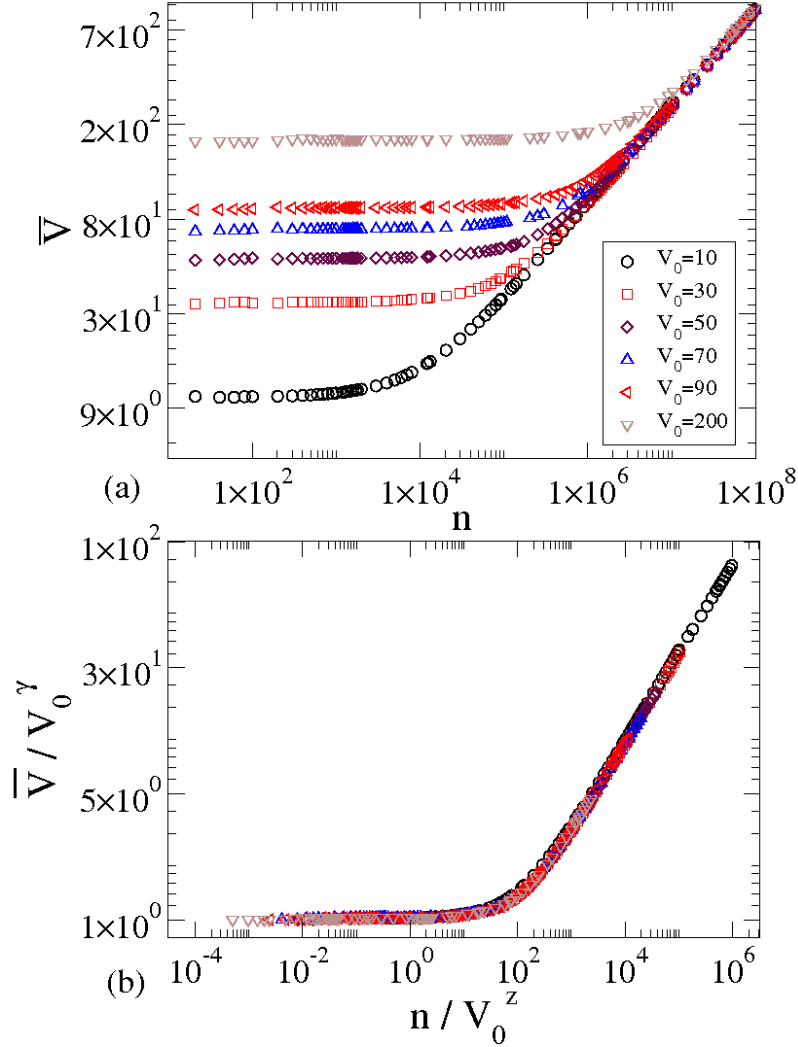}}
\caption{{(a) Behaviour of the average velocity for different values
of $V_0$, a subset of the curves of figure \ref{fig20a}; (b) their collapse onto a single and universal plot.}}
\label{fig21}
\end{figure}

\subsection{Numerical results}

Our numerical results for the time-dependent driven elliptical billiard discuss basically
the behaviour of the average velocity of the particle as a function of the number of bounces\footnote{It should be noted that as long as the dynamics is chaotic there is not fundamental difference 
in choosing the discrete time $n$ or the continuous time $t$. Namely, if $\overline{V} \propto n^\beta$, then $\overline{V} \propto t^{\beta/(1-\beta)}$ that is to say the acceleration exponent changes $\beta \rightarrow \beta/(1-\beta)$.} $n$. 
This is done because in a 4D phase space it is difficult or even meaningless to consider individual pointwise initial conditions. 
We obtain the average velocity in two steps. Firstly, we evaluate the average velocity over the orbit for a single initial condition
\begin{eqnarray}
{V}_i={{1}\over{n+1}}\sum_{j=0}^nV_{i,j}~,
\label{eq015}
\end{eqnarray}
where the index $i$ corresponds to a member of an ensemble of initial conditions. Second, we take the average over the ensemble of initial conditions, so that the average velocity is defined as
\begin{eqnarray}
\overline{V}={{1}\over{M}}\sum_{i=1}^MV_i~,
\label{eq16}
\end{eqnarray}
where $M$ denotes the number of different initial conditions. We have
considered $M=200$ in our simulations and fixed the control parameters as $A_0=2$, $B_0=1$. We will concentrate on the influence
of the initial velocity $V_0$ on the behaviour of the average velocity. Our main goal is to describe such a behaviour using scaling arguments.
Figure \ref{fig19} shows the behavior of the average velocity as a function of the number of collisions for different initial conditions and different values of the control parameter $C$.
Figure \ref{fig20a} shows
the behavior of the average velocity $\bar V$ as a function of the number of collisions $n$ for different initial velocities and fixed $C=0.2$. We
have chosen $11$ different values for initial velocity $V_0$ while a random choice for
the other variables was made as $t \in [0,2\pi]$, $\theta \in [0,2\pi]$
and $\alpha \in [0,\pi]$. As
one can see, all curves of the $\bar V$ behave quite similarly in the
sense that: (a) for short $n$, up to $n \approx n_x$ the average velocity remains constant equal to $\bar V_{ip}$, where $ip$ means initial plateau,
 and (b) after a crossover for $n >> n_x$, all the curves start to grow with the same exponent. For such a behaviour we propose the following hypotheses, which 
turn out to describe the empirical facts:

\begin{enumerate}
\item{For short $n$, say $n\ll{n_x}$, $\bar V$ behaves according to
\begin{equation}
\bar{V}_{ip}\propto V_{0}^{\gamma}~,
\label{eq613}
\end{equation}
and if the initial plateau is well defined we of course expect $\gamma=1$.
}
\item{For $n \gg n_x$, the average velocity is given by
\begin{equation}
\bar{V}\propto {n}^{\beta}~,
\label{eq614}
\end{equation}
where the exponents $\gamma$ and $\beta$ are critical exponents, namely the initial and acceleration exponents, respectively.
}
\item{The crossover iteration number $n_x$ that marks the change from constant
velocity at the initial plateau to the growth is written as
\begin{equation}
n_x\propto V_0^{z}~,
\label{eq615}
\end{equation}
where $z$ is the third exponent, called the crossover exponent.
}
\end{enumerate}
With these three assumptions, from the method of \cite{refLMS}, we suppose that the average
velocity is described in terms of a scaling function of the type
\begin{equation}
\bar{V}(V_0,n)=\lambda\bar{V}(\lambda^a{V_0},\lambda^b{n})~,
\label{eq616}
\end{equation}
where $\lambda$ is the scaling factor, $a$ and $b$ are scaling
exponents. If we chose $\lambda^{a}V_0=1$, then $\lambda=V_0^{-1/a}$ and Eq. (\ref{eq616}) is given by:
\begin{equation}
\bar{V}(V_0,n)=V_0^{-1/a}\bar{V}_1(V_0^{-b/a}n)~,
\label{eq617}
\end{equation}
where $\bar{V}_1(V_0^{-b/a}n)=\bar{V}(1,V_0^{-b/a}n)$ is assumed to be constant for  $n\ll{n_x}$. 
Comparing Eq. (\ref{eq617}) and Eq. (\ref{eq613}), we obtain $\gamma=-1/a$.
On the other hand, if we chose $\lambda^bn=1$, which means $\lambda=n^{-1/b}$ and Eq. (\ref{eq616}) is rewritten as
\begin{equation}
\bar{V}(V_0,n)=n^{-1/b}\bar{V}_2(n^{-a/b}V_0)~,
\label{eq618}
\end{equation}
where the function $\bar{V}_2$ is defined as $\bar{V}_2(n^{-a/b}V_0)=\bar{V}(n^{-a/b}V_0,1)$. It is assumed to be constant for $n\gg{n_x}$. 
Comparing Eq. (\ref{eq618}) and Eq. (\ref{eq614}) we find $\beta=-1/b$. Considering the two different expressions for the 
scaling factor $\lambda$, the condition of the crossover exponent is $\lambda=n_x^{-1/b}=n_x^{\beta}=V_0^{-1/a}=V_0^{\gamma}$, and we find  

\begin{equation}
z={\gamma\over\beta}~.
\label{eq619}
\end{equation}

Note that the scaling exponents are determined if the critical
exponents $\gamma$ and $\beta$ are numerically obtained. The exponent
$\beta$ is obtained from a power law fitting for the average velocity
when $n\gg{n_x}$. Thus, an average of these values gives
$\beta=0.50(1)$ (see Fig. \ref{fig20a}), and $\beta=0.49970(2)$ for others values of $C$ (Fig. \ref{fig19}).  Figure \ref{fig20} shows the behaviour of (a),
${\bar V}_{ip} vs. V_0$ and (b), $n_x vs. V_0$. From a power
law fitting we obtain $z=2.0123(1) \cong 2$ and $\gamma=0.99991(8)$.
Considering the previous values of both $\beta$ and $\gamma$  and using
Eq. \ref{eq619}, we find that $z=1.99(4)$. This result indeed
agrees with our numerical data.
A confirmation of the initial hypotheses comes from a collapse of different curves of ${\bar V} vs. n$ onto a single and universal plot,
as demonstrated in Fig. \ref{fig21}. Additionally, considering the fact that the critical exponents are $\beta \cong 0.5$, $\gamma \cong 1$ and $z \cong 2$ we can conclude 
that the conservative time 
dependent elliptical billiard belongs to the same class of universality like the conservative time-dependent Lorentz gas \cite{ref13a} for the range of the control parameters considered.

\section{Final Remarks}
\label{sec3}

As a conclusion of the present paper, we have studied 
some dynamical properties of a time-dependent elliptical billiard considering elastic collisions with the boundary.
We have shown that the average velocity for short time remains constant but after a crossover it starts to grow with exponent $0.5$. 
We have shown that such a behaviour can be described using scaling properties and we have found an analytical relation between the critical
exponents $\beta$, $\gamma$, and $z$ (acceleration, initial and crossover exponents, respectively). Our scaling hypotheses are confirmed by a 
good collapse of all the curves of the average velocity onto a single universal plot, therefore confirming that the model is scaling invariant and also 
that it belongs to the same class of universality of the time-dependent Lorentz gas for the range of the control parameters considered.
It should be noted that the physical origin and explanation of the scaling laws discovered by \cite{refLMS} is still largely unknown, although there is some progress on the theoretical side \cite{refbr}.

\section*{Acknowledgments}
It is our pleasure to dedicate this work to our friend Professor Tassos Bountis on the occasion of his 60th birthday.
D.F.M.O gratefully acknowledges the financial support by the Slovenian Human Resources Development and Scholarship Fund (Ad futura Foundation). 
 M. R. acknowledges the financial support by The Slovenian Research Agency (ARRS).

%\end{multicols}
\end{document}